\documentclass{epl}
\usepackage{amssymb}

\institute{                    
  \inst{1} Theoretische Physik, ETH-Hönggerberg, CH-8093 Zürich, Switzerland\\
  \inst{2} Centre de Physique Moléculaire Optique et Hertzienne, Université Bordeaux 1-UMR 5798, CNRS, F-33405 Talence Cedex, France\\
\inst{3} Institut Universitaire de France, Paris, France\\
}
\pacs{74.25.q}{Properties of type I and type II superconductors }
\pacs{74.25.Qt }{Vortex lattices, flux pinning, flux creep}
\pacs{42.62.Fi}{Laser spectroscopy }
\input{tcilatex}

\begin{document}

\title{Single Molecule Spectroscopy as a possible tool to study the electric
field in superconductors}
\author{M. Faur\'{e}%
\inst{1,}
\and B. Lounis%
\inst{2,3}
\and A. I. Buzdin%
\inst{2,3}%
}
\maketitle

\begin{abstract}
We discuss the influence of the superconducting transition\ in a film on the
fluorescence spectrum of \textbf{a} single molecule located nearby. We show
that single molecule spectroscopy (SMS) should be an appropriate tool to
detect the electric charge associated with a vortex line. Also we argue that 
\textbf{an} electric field must appear at the boundaries between normal and
superconducting regions of the intermediate phase and SMS could be useful to
study this effect.
\end{abstract}

\section{Introduction}

The studies of the electric field effects in superconductors have a rather
long history (see Ref.\ \cite{Lipavskii02} and articles cited there). In
spite of a large number of theoretical works there exist only a few
experimental hints on the experimental observation of electric field in
superconductors. A regain of interest to this subject has been stimulated by
the prediction of the existence of an electric charge of the vortex core
(see Ref. \cite{Khomskii}). It was argued there that the vortex charge can
be responsible for the sign change in the Hall effect below $T_{c}$.
Subsequent studies \cite{Blatter} stressed the importance of the metallic
screening which strongly reduces the vortex core charge. For the moment,
there is no direct experimental evidence of the charge of a single vortex.
Note that high-resolution nuclear quadrupole frequency measurements revealed
a change in the local electric field at the transition into the mixed state
in YBaCuO \cite{Matsuda}. However, this behavior can also be explained by a
magnetostriction effect due to the coupling between the vortex lattice and
the crystal structure. Therefore, it remains of considerable interest to
directly observe the charge of a single vortex.

Another interesting example of electric field generation in a superconductor
is the intermediate state in type I material. This state involves a sequence
of normal and superconducting regions and appears for example in
superconducting films submitted to a perpendicular magnetic field. The
electric charge in this case appears at the boundaries between the normal
and superconducting phases. The electric field in the intermediate state
outside the superconductor decreases more slowly with the distance from the
surface compared to the vortex field, which can facilitate it registration.

In this paper, we will show that single-molecule spectroscopy \cite{Basche}
should provide a very sensitive probe for the charge distribution and
electric fields in superconductors. Free from ensemble averaging, small
shifts of the narrow single molecule optical lines induced by an external
perturbation can be measured with high accuracy at liquid Helium
temperatures. This technique has been successfully used to probe the
dynamics of low-energy excitations (two-level systems) in polymers \cite%
{Zumbusch} and to reveal interesting phenomena in the electric conduction of
semi-conductors \cite{caruge}. Therefore it may be quite appropriate to
perform the challenging experiments on the observation of the electric
charge of a single vortex or of the domain wall charge in the intermediate
superconducting state.

Single molecule lines being unaffected by the modest magnetic field of the
vortex, the SMS will thus selectively feel the electric charge associated to
the vortex and will be complementary to the decoration and STM techniques of
vortex imaging.

\section{Single Molecule Spectroscopy}

Single molecule spectroscopy is based on the detection of the laser-induced
fluorescence from a small volume of a solid matrix doped with highly
fluorescent molecules, in which at most one and only one chromophore
molecule can be excited by the laser \cite{orrit}. As the red-shifted
fluorescence photons can be efficiently separated from stray laser photons
scattered from the intense exciting beam, the weak single molecule signal
appears on a very low background, with a high signal-to-noise ratio. At
liquid helium temperatures, the zero-phonon lines of the electronic
component (0-0 vibronic line) of a single molecule can become extremely
narrow, while the center frequencies of different molecules are still spread
over a broad inhomogenous band \cite{Tamarat}. The inhomogeneous broadening
arises from the many defects in the solid matrix, which shift
single-molecule lines at random. Therefore, for each particular laser
frequency, resonance is achieved only for a very small fraction of the
molecules in the sample. In addition to the spatial selection (use of small
excitation volumes of tightly focused beams), the laser thus selects
molecules spectrally too.

At low temperatures the dephasing of the optical transition dipole due to
phonons vanishes. For few well chosen fluorophores-matrix systems, such as
the polycyclic aromatic chromophores (dibenzo-anthanthrene or Terrylenne)
embedded in a molecular crystal (Naphtalene,..) or n-Hexadecane Shpol'skii
matrix, the narrow zero-phonon line has a spectral width limited by the
lifetime of the molecule excited state ($\sim $10-40 MHz). This allows one
to employ single chromophore molecules as highly sensitive probes for their
local nano-environments. Internal processes, and external perturbations,
which shift the resonance frequency of a single molecule on the order of 10$%
^{-7}$ of the absolute optical frequency, are easily detectable.

An external electric field changes the spectral position of the molecular
transition by:

\begin{equation}
\Delta \nu =-\frac{1}{\hbar }\left( \delta \overrightarrow{p}\cdot 
\overrightarrow{E}_{loc}+\frac{1}{2}\overrightarrow{E}_{loc}\cdot \delta 
\overleftrightarrow{\alpha }\cdot \overrightarrow{E}_{loc}\right) ,
\end{equation}%
where $\delta \overrightarrow{p}$ is the difference between the dipole
moment vectors and $\delta \overleftrightarrow{\alpha }$ the difference
between the tensors of electrical polarizability between the excited and
ground state of the molecule. $\overrightarrow{E}_{loc}$ is the local
electrical field. Centrosymmetric molecules such as dibenzo-anthanthrene and
terrylene exhibit only a quadratic Stark effect in a vacuum, but when
embedded in a solid matrix, they usually gain a permanent dipole moment due
to distortions by the surrounding matrix. This leads to an additional linear
contribution to the Stark shift, which is usually much stronger than the
quadratic shift. In strongly disordered matrices such as polymers\textbf{,} $%
\delta \overrightarrow{p}$ can be as large as 1$D$ \textbf{(}Debye\textbf{)}%
, but it is around 0.3$D$ in n-Hexadecane Shpol'skii matrix (corresponding
to the frequency shift $\sim $2 MHz/(kV/m)) \cite{Brunel} . Thus local
electric fields of $\sim $10 kV/m will induce a shift of a single
dibenzo-anthanthrene line of the order of its width ($\sim $20 MHz).

Zeeman shift of the singlet-singlet transition of planar aromatic molecule
line is extremely small \cite{Bauer}. Since the molecule has no permanent
magnetic moment, the shift is quadratic and is less than -80 MHz/T$^{2}$.
This means that single molecules line are unaffected by moderate magnetic
fields such as that created by vortex in its neighborhood.

\section{Detection of a single vortex charge by single molecule spectroscopy}

In this section, we briefly remind of the mechanism of the generation of the
charge associated with a vortex line (for more details, see Ref. \cite%
{Blatter}). The geometry of the studied system is displayed in Fig. 1. The
molecule is supposed to be at the distance $z_{0}$ from the superconducting
film (because of the presence of the solid matrix).

The transition into a superconducting state changes the energy spectrum near
the Fermi-level and leads to a variation of the chemical potential $\mu $ as 
$\delta \mu \sim -\Delta ^{2}/\mu _{0\text{ }}$(see \cite{Khomskii}) where $%
\Delta $ is the superconducting energy gap and $\mu _{0\text{ }}$ is the
chemical potential in the normal state. The particle-hole asymmetry also
gives a contribution to $\delta \mu $ of the same order of magnitude \cite%
{Blatter}. We\ consider below the vortices in type II superconductors and
the intermediate state in type I superconductors. In both cases the system
comprises a normal (the vortex core for the type II superconductor) and a
superconducting parts. The equality of the electrochemical potentials leads
to the electrical charge redistribution between normal and superconducting
regions. Indeed, the energy gain due to the condensation of pairs lowers the
energy of the charge carriers and then provokes some carrier influx from the
normal region.

Following Ref. \cite{Blatter} the spatial variation of the gap near the
vortex core can be modelled by $\Delta ^{2}(\rho )=\Delta _{0}^{2}\rho
^{2}/(\rho ^{2}+\xi ^{2})$. Then, the charge variation will be (for $\rho
\ll \lambda $, where $\lambda $ is the penetration depth)

\begin{equation}
\delta q_{ext}=\frac{KeN(\mu _{0\text{ }})\Delta _{0}^{2}\rho ^{2}}{\mu _{0%
\text{ }}\left( \rho ^{2}+\xi ^{2}\right) },  \label{delta n}
\end{equation}%
where the coefficient $K$ is of the order of unity and depends on the
details of the electronic band structure. However, this charge
redistribution is strongly screened\textbf{;} in the Thomas-Fermi
approximation the electrostatic potential $\varphi (\overrightarrow{r})$ in
the superconductor may be determined from

\begin{equation}
(\nabla ^{2}-\frac{1}{\lambda _{TF}^{2}})\varphi (\overrightarrow{r})=-4\pi
\delta q_{ext}.  \label{Thomas-fermi}
\end{equation}

In the limit $\xi \gg \lambda _{TF}$ the potential in the first
approximation is $\varphi (\overrightarrow{r})=4\pi \lambda _{TF}^{2}\delta
q_{ext}$ and the resulting charge density $\delta q$ may be obtained from
the Poisson equation $\nabla ^{2}\varphi (\overrightarrow{r})=-4\pi \delta q$%
. We see that the electrostatic screening strongly reduces the "bare" charge
redistribution and may be determined from $\delta q=-\lambda _{TF}^{2}\nabla
^{2}\left( \delta q_{ext}\right) $.

Outside the sample the electrostatic potential is maximal on the vortex axis
and given by

\begin{equation}
\varphi _{out}=-\frac{8\pi K\lambda _{TF}^{2}\xi ^{2}eN(\mu _{0\text{ }%
})\Delta _{0}^{2}}{\mu _{0\text{ }}}\dint\limits_{0}^{\infty
}\dint\limits_{0}^{\infty }dz\rho d\rho \frac{\xi ^{2}-\rho ^{2}}{\left(
\rho ^{2}+\xi ^{2}\right) ^{3}\sqrt{\rho ^{2}+(z+z_{0})^{2}}},
\label{potential}
\end{equation}%
where $z_{0}$ is the coordinate of the molecule on the $z$ axis. The
electrical field can then be deduced from the above expression. The
component of the electric field along the $z$ axis denoted $E_{z}$ is%
\begin{eqnarray}
E_{z} &=&-\frac{\partial \varphi _{out}}{\partial z_{0}} \\
&=&\frac{4\pi K\lambda _{TF}^{2}eN(\mu _{0\text{ }})\Delta _{0}^{2}}{\mu
_{0}\xi }\dint\limits_{0}^{\infty }du\frac{u-1}{\left( u+1\right) ^{3}\sqrt{%
u+a}}.  \label{electric_fieldint}
\end{eqnarray}%
where $u=\left( \rho /\xi \right) ^{2}$ and $a=\left( z_{0}/\xi \right) ^{2}$%
. In the case $z_{0}>\xi $ ($a>1)$, $E_{z}$ becomes

\begin{equation}
E_{z}=\frac{4\pi K\lambda _{TF}^{2}eN(\mu _{0\text{ }})\Delta _{0}^{2}}{\mu
_{0}\xi }\left[ \frac{3\sqrt{a}}{2(a-1)^{2}}-\frac{(2a+1)}{2(a-1)^{5/2}}\ln (%
\sqrt{a}+\sqrt{a-1})\right] .  \label{electric_field}
\end{equation}
Note that in the case $a<1$, the expression of $E_{z}$ is a bit more
complicated, and therefore not presented here.

When the molecule is far from the vortex (that is $z_{0}\gg \xi $), $E_{z}$
can be simplified into%
\begin{equation}
E_{z}\thickapprox -\frac{4\pi K\lambda _{TF}^{2}eN(\mu _{0\text{ }})\Delta
_{0}^{2}}{\mu _{0}\xi }\left[ \left( \frac{\xi }{z_{0}}\right) ^{3}\ln \frac{%
z_{0}}{\xi }\right] ,  \label{vortex field}
\end{equation}%
which corresponds to the asymptotic formula obtained in Ref. \cite{Blatter}.
Note that this formula is valid for $z_{0}<\lambda $, otherwise $z_{0}$ has
to be replaced by $\lambda $ in the logarithm term in (\ref{vortex field}).
The electric field at the surface of superconductor ($z_{0}=0$) as it
readily follows from expression (\ref{electric_fieldint}) is%
\begin{equation}
E_{z}(0)=-\frac{\pi ^{2}K\lambda _{TF}^{2}eN(\mu _{0\text{ }})\Delta _{0}^{2}%
}{\mu _{0\text{ }}\xi }.  \label{field_surface}
\end{equation}

At the surface of the superconductor, the electric field created by the
charge associated to the vortex core (\ref{electric_fieldint}) depends on
the temperature following the ratio $\Delta _{0}^{2}/\xi $. More precisely,
since $\Delta _{0}^{2}\sim (T_{c}-T)$ and $\xi \sim (T_{c}-T)^{-1/2}$, the
evolution of $E_{z}$ as a function of temperature is given by $%
(T_{c}-T)^{3/2}$. Therefore, the electric field effect quickly disappears
when the temperature approaches the critical temperature. In other words,
the experiment should better be performed at sufficiently low temperature $%
\left( T\lesssim 0.5T_{c}\right) $. At large distances, $z_{0}\gg \xi $, the
temperature dependence is given by $\Delta _{0}^{2}\xi ^{2}$. However, this
coefficient does not depend on the temperature, and, as a result, so does $%
E_{z}$. Therefore, the variation of the electric field as a function of the
temperature depends on the thickness of the solid matrix containing the
studied molecule.

One can notice from expression (\ref{vortex field}) that the vortex electric
field, and as a result the detected effect, rapidly decreases as a function
of the distance between the molecule and the vortex ($\sim z_{0}^{-3}$). The
thickness of the solid matrix should therefore be as thin as possible. One
can roughly estimate the value of the electric field at the surface of the
superconducting vortex considering expression (\ref{field_surface}): $%
E_{z}(0)\sim e/\left( \xi a_{B}\right) (T_{c}/\mu _{0\text{ }})^{2}.$The
parameter $a_{B}$ is the Bohr radius and is about $0.1$ nm. If the
superconducting coherence length $\xi $ is $10$ nm and $T_{c}/\mu _{0\text{ }%
}\sim a_{B}/\xi $,\ then\ $E_{z}(0)\sim ea_{B}/\left( \xi ^{3}\right) \sim
10^{5}$ V/m, which is of the same order of magnitude as the SMS sensitivity.

From an experimental point of view it is unlikely that the fluorescent
molecule will be just above the vortex. It may be then more appropriate to
change the external magnetic field, which would make the vortices move. In
such a setup is possible to detect the movement of the vortex when it passes
below the SMS molecule. However the distance $z_{0}$ between the fluorescent
molecule and superconductor needs to be smaller than the period $a_{A}$ of
the Abrikosov lattice. For $z_{0}>a_{A}$ the amplitude of the electric field
modulation decreases exponentially as $exp(-2\pi z_{0}/a_{A})$ \cite{Blatter}%
.

\section{Intermediate state}

The system studied in this section is a film in the so-called intermediate
state, as shown in Fig. 2. The intermediate state appears in type I
superconducting films when a perpendicular magnetic field is applied. In
this state, regions of normal and superconducting Meissner phases alternate.
The $x$ axis is chosen perpendicular to the S/N boundary (located at $x=0$).
The N state occupies the space where $x<0$ while the S state corresponds to $%
x>0$.

In the intermediate state, the form of the order parameter variation can be
extrapolated from the Ginzburg Landau theory \cite{DeGen} to the low
temperatures as $\Delta (x)=\Delta _{0}\tanh \left( x/\sqrt{2}\xi \right) $.
Taking into account the metallic screening, the charge density being%
\begin{equation}
\delta q=\frac{K\lambda _{TF}^{2}eN(\mu _{0\text{ }})\Delta _{0}^{2}}{\mu _{0%
\text{ }}\xi ^{2}}\left( \frac{3}{\cosh ^{4}\left( x/\sqrt{2}\xi \right) }-%
\frac{2}{\cosh ^{2}\left( x/\sqrt{2}\xi \right) }\right) .
\label{density_charge_intermediate}
\end{equation}

As in the previous section, the electric field created above the
superconductor can be deduced from the expression of the potential resulting
from the density charge (\ref{density_charge_intermediate}). If the molecule
is located at the coordinates $x_{0}$ and $z_{0}$, the components of the
electric field may be written as

\begin{equation}
E_{z}=\frac{K\lambda _{TF}^{2}eN(\mu _{0\text{ }})\Delta _{0}^{2}}{\mu _{0%
\text{ }}\xi ^{2}}\dint\limits_{0}^{\infty }\dint\limits_{-\infty }^{\infty
}\dint\limits_{0}^{\infty }\frac{dxdydz(z+z_{0})}{\left( \left(
x-x_{0}\right) ^{2}+(z+z_{0})^{2}+y^{2}\right) ^{3/2}}\left( \frac{3}{\cosh
^{4}\frac{x}{\sqrt{2}\xi }}-\frac{2}{\cosh ^{2}\frac{x}{\sqrt{2}\xi }}%
\right) ,
\end{equation}%
\begin{equation}
E_{x}=\frac{K\lambda _{TF}^{2}eN(\mu _{0\text{ }})\Delta _{0}^{2}}{\mu _{0%
\text{ }}\xi ^{2}}\dint\limits_{0}^{\infty }\dint\limits_{-\infty }^{\infty
}\dint\limits_{0}^{\infty }\frac{dxdydz\left( x_{0}-x\right) }{\left( \left(
x-x_{0}\right) ^{2}+(z+z_{0})^{2}+y^{2}\right) ^{3/2}}\left( \frac{3}{\cosh
^{4}\frac{x}{\sqrt{2}\xi }}-\frac{2}{\cosh ^{2}\frac{x}{\sqrt{2}\xi }}\right)
\end{equation}%
At large distances $\frac{z_{0}}{\sqrt{2}\xi }>>1$, we have 
\begin{equation}
E_{z}\approx -\frac{\lambda _{TF}^{2}eKN(\mu _{0\text{ }})\Delta _{0}^{2}}{%
\mu _{0\text{ }}\xi }\sqrt{2}\frac{\widetilde{x}_{0}-1}{\left( \widetilde{z}%
_{0}\right) ^{2}},
\end{equation}%
where $\widetilde{x}_{0}=\frac{x_{0}}{\sqrt{2}\xi }$. The decrease of $E_{z}$
in the intermediate state is slower than in the vortex case $(\sim 
\widetilde{z}_{0}^{-3})$.

The component $E_{x}$ of the electric field parallel to the surface
decreases even more slowly, as $E_{x}\sim \widetilde{z}_{0}^{-1}$. Indeed,
at large distances $\widetilde{z}_{0}>>1$, it can be written as

\begin{equation}
E_{x}\approx -\frac{\sqrt{2}\lambda _{TF}^{2}KeN(\mu _{0\text{ }})\Delta
_{0}^{2}}{\mu _{0\text{ }}}\frac{1}{z_{0}}.
\end{equation}%
Note that $E_{x}$ depends on the temperature as $T_{c}-T$. At the surface of
the superconductor, if the molecule is above the N region ($\widetilde{x}%
_{0}<0$), $E_{x}=0$ while if it is above the S state ($\widetilde{x}_{0}>0$%
), $E_{x}$ can be expressed as%
\begin{equation}
E_{x}\approx \frac{\lambda _{TF}^{2}KeN(\mu _{0\text{ }})\Delta _{0}^{2}}{%
\mu _{0}\xi }8\sqrt{2}\pi \frac{e^{2\widetilde{x}_{0}}\left( e^{2\widetilde{x%
}_{0}}-1\right) }{\left( e^{2\widetilde{x}_{0}}+1\right) ^{3}}.
\end{equation}%
Note that the maximum field is achieved at $\widetilde{x}_{0}=\ln \left( 2+%
\sqrt{3}\right) /2\approx 0.66$ and is equal to $E_{x}^{\max }=\frac{\lambda
_{TF}^{2}eKN(\mu _{0\text{ }})\Delta _{0}^{2}}{\mu _{0}\xi }c$ ($c=4\pi 
\sqrt{2}/3\sqrt{3}$). In this case, the temperature dependence is $\left(
T_{c}-T\right) ^{3/2}$. An estimate of an electric field above the S/N
interface gives the field of same order of magnitude as above the vortex
core. Therefore the SMS could detect the electric field in the intermediate
state too.

\section{Conclusion}

We have argued that single molecule spectroscopy may be a useful tool to
study the electric field effects in superconductors. The order of magnitude
of the electric field above the vortex core or near the S/N interface in the
intermediate state is about $10^{5}$ V/m for a distance of $\sim 10$ nm, is
one order of magnitude larger than the limit of electric field detection by
Stark-shift of single molecule lines. Because of metal quenching \cite%
{Barnes},\cite{Milonni} (which induces fluorescence intensity decrease and
line-broadening), molecule-superconductor-substrate distances should be
larger than 20 nm. At these distances, charge distributions at the S/N
interface are still detectable (because of the 1/$z_{0}$ decrease of the
field). However for vortices one can rapidly reach the sensitivity limit of
electric field detection (due to $1/z_{0}^{3}$ decrease of the field).

To improve the sensitivity, one can use very thin layers of superconductors
to limit the metal quenching effects. Another possibility would be to apply
a time modulated magnetic field or electric current to the superconductor.
This will induce an oscillation of the molecule-vortex (or S/N interface)
distance and thus a modulation of the single molecule resonance position.
When the molecule is excited at a fixed laser wavelength, detecting the
fluorescence intensity fluctuations at the magnetic field (electric current)
modulation frequency can improve the sensitivity by more than one order of
magnitude and thus ensure an accurate estimation of the electric field
effects.

Among the perspectives of the SMS method, we should mention the study of the
dynamics of the vortex flow and S/N boundary displacements in the
intermediate phase. Also if the SMS sensitivity could be increased, this
method could be applied to study such a subtle effect as the Bernoulli
potential in the Meissner state (see for example \cite{Lipavskii04}).

Finally note that even in the absence of vortices or intermediate state the
superconducting transition may influence SMS spectra. Indeed the transition
into the superconducting state leads to a change of the permittivity \cite%
{abrikosov} $\delta \epsilon /\epsilon \sim \Delta ^{2}/\omega ^{2}\ln
\left( \omega /\Delta \right) $, in the limit $\omega \gg \Delta $, as it is
realized in the experimental situation. The modification of the permittivity
leads to a shift in the fluorescence frequency due to dipole mirror image
charge effect. The change of the frequency is however quite small and may be
of some relevance only for high-T$_{c}$ superconductors or in the case of
the fluorescence in the near infrared. One may also expect that the
superconducting transition could significantly modify the decay rate of the
single molecule fluorescence.

\acknowledgments Acknowledgements. The authors thank M. Orrit, G. Blatter,
H. Courtois, D. Geshkenbein, P. Lipavsky, D. Ryzhov and P. Tamarat for
useful discussions. This work was supported in part by EGIDE Program
10197RC. M. F. acknowledges the support of the Swiss NCCR program MaNEP.

\end{document}